\documentclass{webofc}

\usepackage[varg]{txfonts}   
\usepackage{hyperref}
\usepackage{url}
\hypersetup{colorlinks=true,citecolor=blue,urlcolor=blue,linkcolor=blue}
\begin{document}
\title{Unexpectedly large violation
of isospin symmetry\\
in
nucleus--nucleus collisions
at CERN SPS energies}
%
%

\author{\firstname{Andrzej} \lastname{Rybicki}\inst{1}\fnsep\thanks{\email{andrzej.rybicki@ifj.edu.pl}}\\
for the NA61/SHINE Collaboration
}

\institute{The Henryk Niewodniczański Institute of Nuclear Physics Polish Academy of Sciences,\\ ul. Radzikowskiego 152,
31-342 Krak\'ow, Poland
          }

\abstract{The NA61/SHINE experiment at the CERN SPS reports an unexpectedly large excess in the inclusive production of charged over neutral kaons in 0--10\% central Ar+Sc collisions at $\sqrt{s_\mathrm{{NN}}}=8.8$ and 11.9 GeV. This effect corresponds to an unexpectedly large violation of isospin symmetry, that cannot be explained by known mechanisms contained in the present theoretical models. The comparison of the NA61/SHINE results to data from other experiments reveals, within sizeable uncertainties, a consistent isospin symmetry-violating excess of charged kaon production in the energy regime $2.6<\sqrt{s_\mathrm{NN}}<200$~GeV. At LHC energies ($0.9<\sqrt{s_\mathrm{NN}}<13$~TeV) the charged-over-neutral kaon ratio appears consistent with unity, but the uncertainties remain sizeable. A similar excess of charged over neutral kaon yields is observed by NA61/SHINE in minimum bias $\pi^-$+C reactions at $\sqrt{s_\mathrm{NN}}=17.3$ and 25.7~GeV, also in disagreement with model predictions. No clear theoretical explanation of these phenomena has been provided so far. An unexpectedly large violation of isospin symmetry brings potential heavy consequences for our understanding of particle production, and calls for a concentrated experimental and theoretical effort.}
\maketitle
\section{Introduction: multiparticle production}
\label{intro}

This paper focuses on hadron--nucleus and nucleus--nucleus collisions at center-of-mass (cms) energies of the order of 10 GeV per nucleon pair. Such processes are governed by (mostly) non-perturbative strong interactions, and result in the production of a large number -- tens, hundreds -- of new particles in the final state (multiparticle production). Isospin symmetry (IS) which can be considered as a remnant of the QCD flavor symmetry between $u$ and $d$ quarks, remains an essential even if approximate feature of the production process. At energies equal (or higher) to those specified above, the bulk features of multiparticle production (inclusive spectra, total multiplicities, etc) are widely believed to obey IS up to a very good accuracy, of the order of 3$-$4\%. Nevertheless, the simultaneous measurement of inclusive spectra and yields of charged ($K^+$\hspace*{-0.1cm}, $K^-$) and neutral ($ K^0_S$) mesons reported by the NA61/SHINE experiment in central Ar+Sc collisions at $\sqrt{s_\mathrm{NN}}=11.9$~GeV~\cite{1} reveals an unexpectedly large IS violation (18\,$\pm$\,6\% at kaon mid-rapidity). As will be discussed below, measurements from other experiments over a wide range of collision energy appear consistent with this observation. At the present moment (Oct.,~2026), the origin of this effect remains unclear.

\section{The measurement}
\label{themes}

The NA61/SHINE experiment is a wide-acceptance, fixed-target spectrometer located at the H2 beam line of the CERN SPS. The details of the detector setup, including among others several large Time Projection Chambers (TPC) and a forward calorimeter for collision centrality triggering and selection, are given in Ref.~\cite{2}. The reconstruction of tracks and measurement of momentum of charged particles is realized by TPC detectors, two of which are located in a magnetic field provided by superconducting magnets. The identification of $K^+$ and $K^-$ mesons proceeds by the measurement of their ionization energy loss (d$E$/d$x$) in the TPCs. Neutral $K^0_S$ mesons are identified by the characteristic $V^0$ topology of the $K^0_S\rightarrow\pi^+\pi^-$~decay. The measured kaon yields are corrected for several detector effects (biases) including the geometrical acceptance, reconstruction efficiency, weak decays, and secondary interactions. The selection of 0--10\% central Ar+Sc collisions is performed through the measurement of the energy deposit in the forward calorimeter. This deposit, predominantly corresponding to the energy of projectile spectator nucleons with the addition of produced charged and neutral particles, is expected to be non-biasing with respect to the charged-over-neutral particle yield ratio, in particular for most central collisions. A detailed discussion of the data analyses of neutral and charged kaon yields, including the analysis of systematic uncertainties which is essential for the present measurement, can be found in Refs.~\cite{1,3}.

\begin{figure*}[!h]
\centering
\vspace*{0.1cm}       
\hspace*{-0.6cm}
\includegraphics[width=0.462\textwidth]{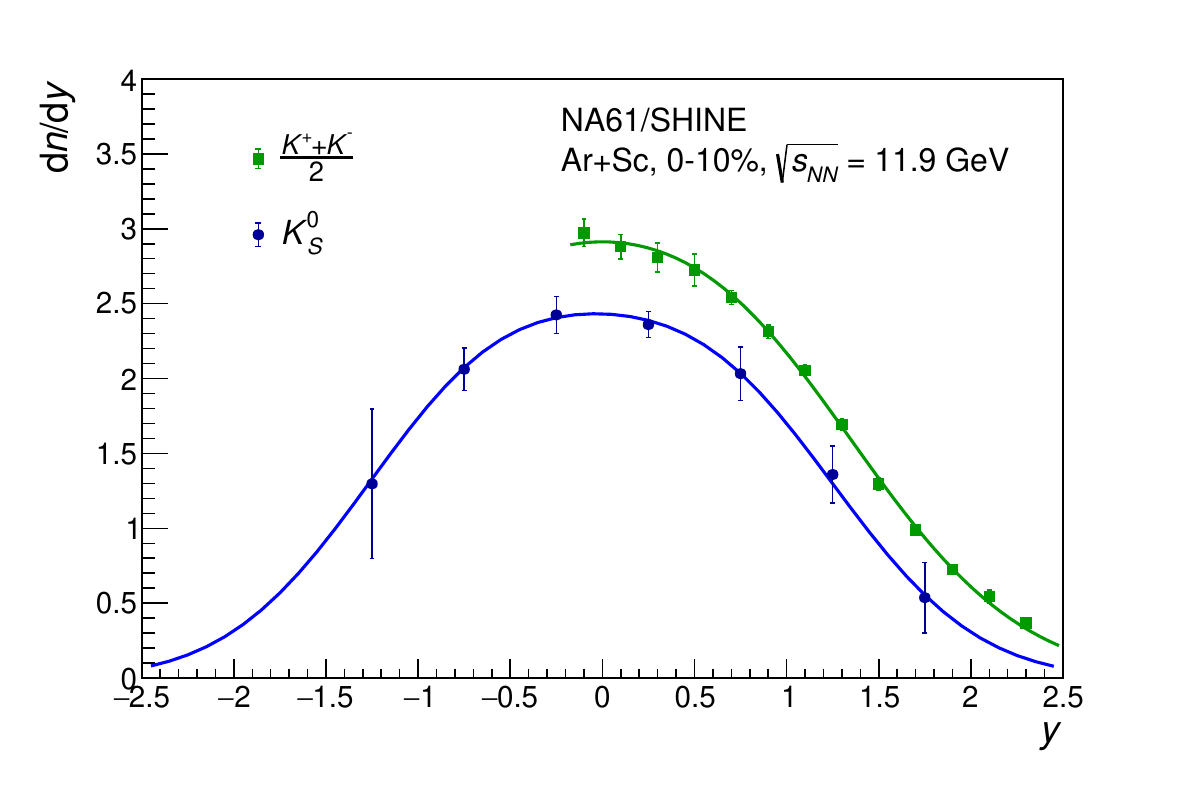}
\hspace*{-0.44cm}
\includegraphics[width=0.583\textwidth]{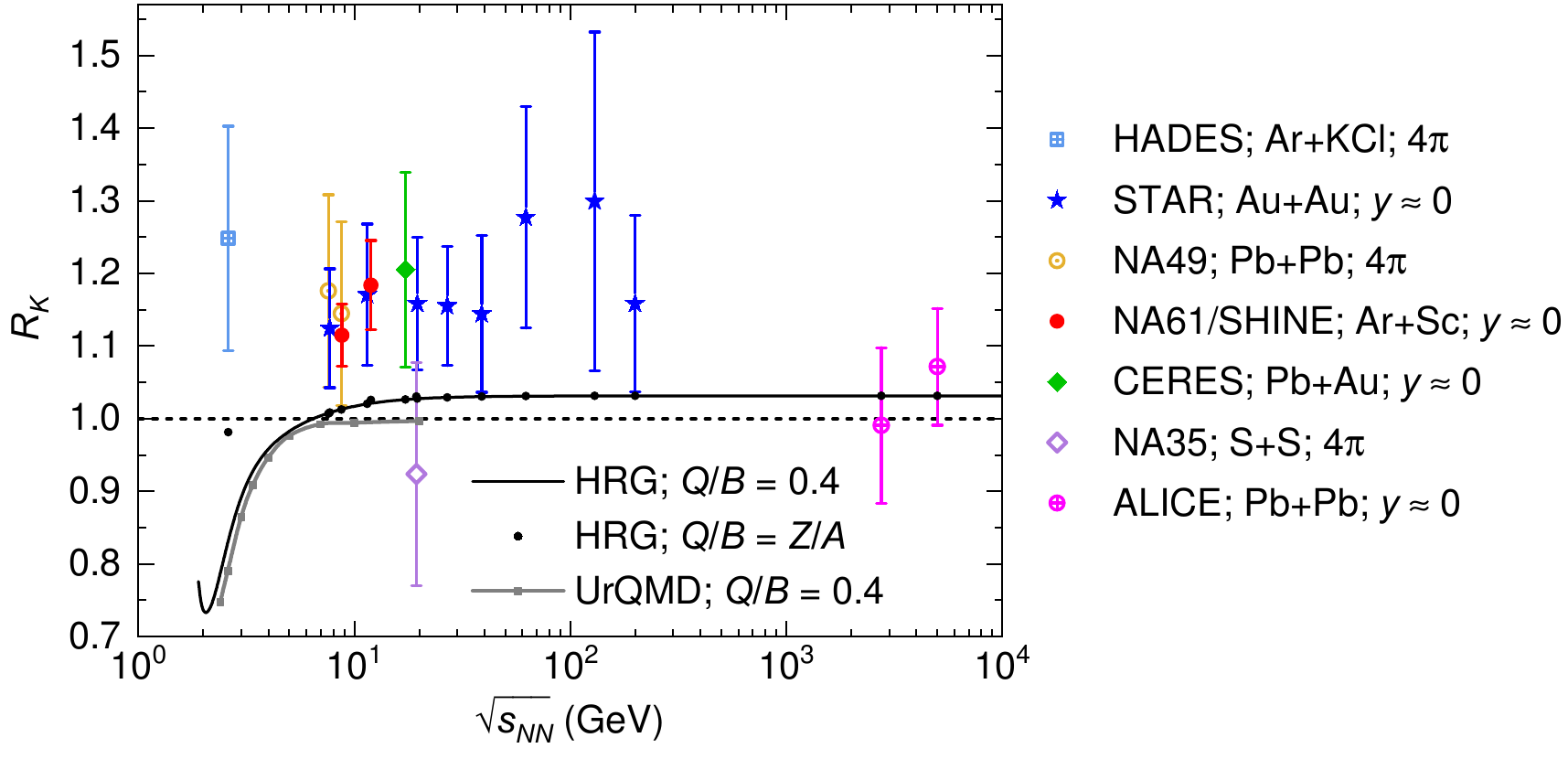}
\hspace*{-0.6cm}
  \begin{picture}(1,1)
 \put(-286,30){\bf(a)}
 \put(-92,81){\bf(b)}
\vspace*{0.3cm}       
  \end{picture}
  \caption{{\bf(a)} Inclusive rapidity spectrum of $K^0_S$ mesons, compared to the average of $K^+$ and $K^-$ rapidity distributions in central Ar+Sc collisions, from Ref.~\cite{1}. 
    {\bf(b)} Ratio $R_K\,$=$\,(\langle K^+\rangle $$+$$\langle K^-\rangle)$/(2$\langle K^0_S\rangle $) drawn as a function of collision energy. The compiled results include ratios of mid-rapidity densities as well as 4$\pi$ yields.  
    HRG and UrQMD model predictions are also shown ($Q/B$ is the total charge-to-baryon ratio assumed in the calculation). From Ref.~\cite{1},
    see therein for the bi\-blio\-graphy,
    updated with 
    the preliminary NA61/SHINE data point at 8.8~GeV~\cite{j} and the ALICE data point at 5.02~TeV~\cite{a7,a8}. Error bars denote total uncertainties.}
\label{f1}       
\vspace*{-0.7cm}       
\end{figure*}

\section{Results and comparison to existing data}
\label{res}

Figure~\ref{f1}(a) presents the inclusive cms rapidity distribution of $K^0_S$ mesons produced in 0--10\% central Ar+Sc collisions, put together with the arithmetic average of $K^+$ and $K^-$ meson distributions produced in the same reactions. The
$^{40}\hspace*{-0.35cm}_{\,\,18\,\,}$\hspace*{-0.08cm}Ar\,+\,\hspace*{-0.01cm}$^{45}\hspace*{-0.35cm}_{\,\,21\,\,}$\hspace*{-0.08cm}Sc
system is nearly charge-symmetric, with the total number of incoming valence $u$ and $d$ quarks equal within ~6\%.\footnote{The
influence of the small excess of neutrons ({\emph i.e.}, of valence $d$ quarks) is included in the predictions of theoretical models discussed below in the text.}
For charge-symmetric systems, the prediction from exact IS writes as follows:
\begin{equation}
\frac{\langle K^+(u\bar{s})\rangle+\langle K^-(\bar{u}s)\rangle}{2}=
%
\frac{\langle K^0(d\bar{s})\rangle+\langle \hspace*{0.03cm}\overline{\hspace*{-0.03cm}K}{\hspace*{0.01cm}^0}(\bar{d}s)\rangle}{2}=
\langle K^0_S\rangle~,
\end{equation}
where the symbols given in the brackets $\langle~\rangle$ correspond to average meson yields per event and the second equality neglects the small effect of CP violation. The experimental result corresponds to an excess of charged kaon over neutral kaon yields, estimated to 18\,$\pm$\,6\% at kaon mid-rapidity ($y=0$). The cited total value of 6\% includes both statistical and (dominant) systematic uncertainties. The cited number corresponds to an unexpectedly large violation of IS in the kaon sector of multiparticle production at SPS energies. Following the estimates given in Ref.~\cite{1}, this effect would result in an average excess of about four (4.2\,$\pm$\,1.6) additional charged $K$ mesons produced per one central Ar+Sc collision if compared to the prediction from exact isospin symmetry.

Figure~\ref{f1}(b) shows the NA61/SHINE result at $\sqrt{s_\mathrm{NN}}=11.9$~GeV, put together with the compilation of results from HADES, STAR, NA49, CERES, NA35, and ALICE experiments. All the results are presented in terms of the ratio 
\begin{equation}
  R_K\equiv
  \frac{\langle K^+\rangle+\langle K^-\rangle}{\langle K^0\rangle+\langle\hspace*{0.03cm}\overline{\hspace*{-0.03cm}K}{\hspace*{0.01cm}^0}\rangle}=
  \frac{\langle K^+\rangle+\langle K^-\rangle}{2 \langle K^0_S\rangle}
\end{equation}
as a function of collision energy. The $R_K$ values in the compilation were obtained from results on $K^+$\hspace*{-0.1cm}, $K^-$\hspace*{-0.1cm}, and $K^0_S$ production released by the respective Collaborations. The details of the procedure are given in Ref.~\cite{1}. The Figure includes a new preliminary NA61/SHINE result from 0--10\% central Ar+Sc collisions at $\sqrt{s_\mathrm{NN}}=8.8$~GeV~\cite{j} which remains consistent with that at 11.9~GeV within uncertainties. In spite of the large error bars apparent for the compiled data points, an overall consistent picture emerges from the Figure.~The \mbox{excess} of charged-over-neutral kaon production remains apparent
in the entire
energy regime \mbox{$2.6<\sqrt{s_\mathrm{NN}}<200$~GeV}. At the higher energies corresponding to the LHC, the values of $R_K$ remain consistent with unity within their sizeable uncertainties. This fact will be commented upon in more detail below (Sec.~\ref{alice}).

In addition to experimental results, the Figure includes predictions from two theoretical models, the Hadron Resonance Gas (HRG) and the microscopic Ultra-Relativistic Quantum Molecular Dynamics (UrQMD) code. The implementation of several known sources of IS violation in the HRG model is detailed in Ref.~\cite{1}. These known sources are not able to explain the observed excess of charged-over-neutral kaon production. 

The cited reference gives an estimate of the significance of the deviation data/theory(HRG) of 4.7$\sigma$. This value is expected to be updated by the addition of new data points including new NA61/SHINE data at other energies, and by the evaluation of possible correlations between the data points coming from a single experiment. The latter correlations were ignored in the present estimate and can be properly evaluated only by the respective Collaborations. The extreme assumption of dominance of systematic uncertainties and a 100\% correlation of systematic uncertainties within each experiment brings the lower limit of the significance data/theory equal to 3.2$\sigma$~\cite{5}. The other extreme assumption of the overestimation of experimental uncertainties, based on the fact that the $\mathrm{\chi^2_{min}}$/{dof} of the deviation data/theory remains significantly below unity for the ensemble of data points shown in Fig.~\ref{f1}(b), results in an upper limit of about 7$\sigma$ (therein). Further work by different experiments is necessary in order to provide a final estimate of the significance. 

\section{Data from ALICE}
\label{alice}

\begin{figure*}[!t]
\centering
\sidecaption
\hspace*{0.3cm}
\includegraphics[trim= 0 -8 132 0,clip,width=0.39\textwidth]{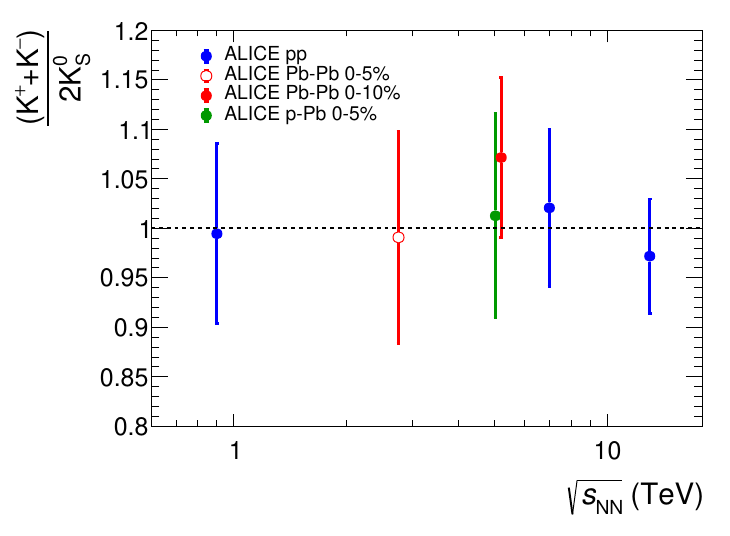}
\hspace*{0.5cm}
\caption{Compilation of results from the ALICE ex\-pe\-ri\-ment in the collision energy regime $0.9$$<$$\sqrt{s_\mathrm{NN}}$$<$$13$ TeV~\cite{a1,a2,a3,a4,a5,a6,a7,a8,a9}. Error bars denote total uncertainties. The Figure comes from Ref.~\cite{5}.}
\label{f3}       
\vspace*{-0.2cm}
\end{figure*}

Figure~\ref{f3} displays a compilation of results from the ALICE experiment in the energy regime $0.9<\sqrt{s_\mathrm{NN}}<13$~TeV. These include Pb+Pb, $p$+Pb, and $p$+$p$ reactions. The obtained values of the charged-over-neutral kaon ratio $R_K$ appear consistent with unity for all the reactions and energies, although the experimental uncertainties of the order of 10\% are too large to allow for definite conclusions. The question whether the IS violation apparent at SIS/SPS/RHIC energies would fade out with increasing \hspace*{-0.05cm}$\sqrt{s_\mathrm{NN}}$ (decreasing baryochemical potential) appears as an issue of great interest, to be solved once more precise experimental data are available.

\section{Excess of charged kaons in small systems}
\label{small}

\begin{figure*}[!h]
\centering
\vspace*{-0.05cm}       
\hspace*{-1cm}
\includegraphics[width=0.52\textwidth]{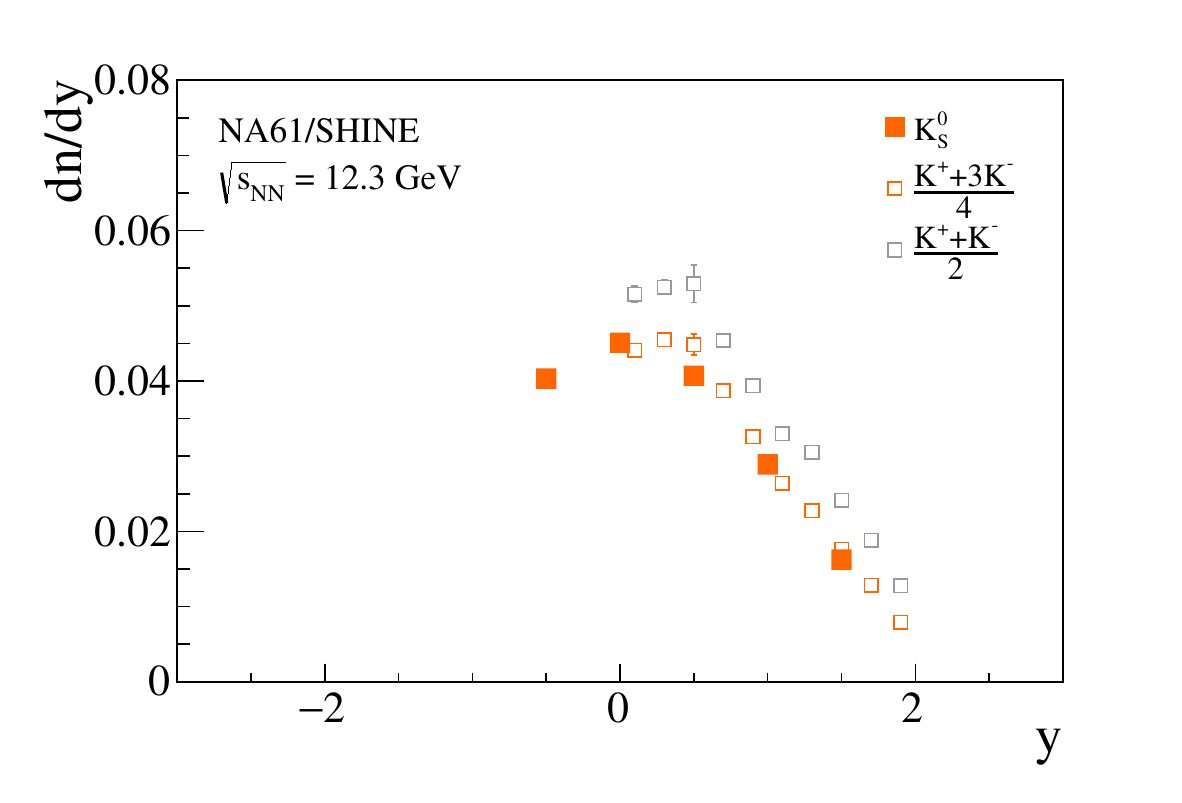}
\includegraphics[trim= 0 -8 175 0,clip,width=0.444\textwidth]{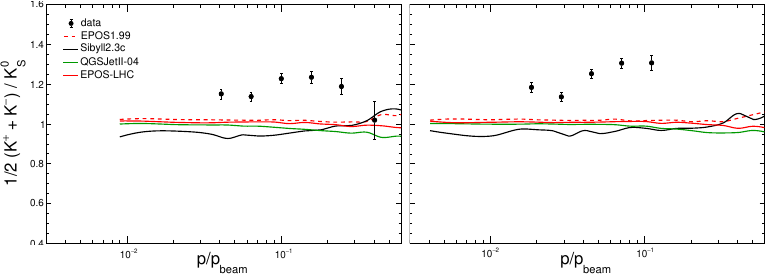}
\hspace*{-0.6cm}
  \begin{picture}(1,1)
\put(-306,34){\bf(a)}
\put(-309,90){\scriptsize{p+p}}
\put(-120,34){\bf(b)}
\put(-39,43){\scriptsize{NA61/SHINE}}
\put(-39,35){\scriptsize{$\pi^-$\hspace*{-0.05cm}+C}}
\put(-41,28){\scriptsize{$\mathrm{\sqrt{s_\mathrm{\pi N}}}$\hspace*{0.03cm}=\hspace*{0.03cm}17.3~GeV}}
 \vspace*{0.3cm}       
  \end{picture}
  \caption{{\bf(a)} Inclusive rapidity spectrum of $K^0_S$ mesons, compared to the average of $K^+$ and $K^-$ rapidity distributions in $p$+$\hspace*{-0.02cm}p$ collisions. The prediction from the Quark Coalescence
    model ($\frac{K^++3K^-}{4}$) is also shown. From Ref.~\cite{k0spp}. 
    {\bf(b)} Ratio $R_K\,$=$\,(\langle K^+\rangle $$+$$\langle K^-\rangle)$/(2$\langle K^0_S\rangle $) drawn as a function $x_F$(lab)$=$$p$/$p$$_\mathrm{beam}$ in $\pi^-$+C reactions, compared to predictions from theoretical models. From Ref.~\cite{m}.}
\label{f5}       
\end{figure*}

The ALICE results from $p$+$\hspace*{-0.02cm}p$ and $p$+Pb collisions can be compared to NA61/SHINE measurements of $p$+$\hspace*{-0.02cm}p$ and hadron--nucleus systems, performed at significantly lower energies. Figure~\ref{f5} shows (a) the comparison of charged and neutral kaon rapidity distributions in $p$+$\hspace*{-0.02cm}p$ collisions, and (b) the $x_F$(lab)-dependence of the $R_K$ ratio in $\pi^-$+C reactions. The presented plots are representative for what is observed at all energies where NA61/SHINE measurements are available ($p$+$\hspace*{-0.02cm}p$:~$\sqrt{s_\mathrm{NN}}=7.7$,~8.8,~12.3,~17.3~GeV, $\pi^-$+C:~$\sqrt{s_{\pi \mathrm{N}}}=17.3$,~25.7~GeV). A charged-over-neutral kaon \mbox{excess} is apparent in all these measurements. Its physical interpretation must take account of the fact that none of the considered colliding systems is charge-symmetric. In $p$+$\hspace*{-0.02cm}p$ collisions, the initial excess of valence $u$ quarks in the proton~($uud$) can enhance the production of charged $K^+(u\bar{s})$ mesons. Indeed, the NA61/SHINE $p$+$\hspace*{-0.02cm}p$ data can be described on that basis by the semi-empirical Quark Coalescence (QC) model preserving~IS~\cite{6,7}. However, the latter model cannot explain the excess observed in $\pi^-$+C reactions, Fig.~\ref{f5}(b). Similarly, all the theoretical models included in the Figure unanimously predict $R_K$ close to unity, and thus fail to describe the experimental data. The latter fact opens the question of top importance of the system size-dependence of IS violation, to be addressed in Secs.~\ref{theory} and \ref{plans} below.

\section{Theory}
\label{theory}

At the present moment (Oct.,~2026), theoretical efforts known to the author provided no clear explanation for the effects apparent in nucleus--nucleus and $\pi^-$+C reactions discussed above. The state-of-the art models QC, PHSD, EPOS, SMASH, UrQMD, and HRG, which contain no new (unknown) sources of IS violation all predict charged-over-neutral kaon ratios in the
overall
range
$0.94\hspace*{-0.1cm}<\hspace*{-0.1cm}R_K\hspace*{-0.1cm}<\hspace*{-0.1cm}1.03$,
in the energy regime $10\hspace*{-0.1cm}<\hspace*{-0.15cm}\sqrt{s_\mathrm{NN}}\hspace*{-0.1cm}<\hspace*{-0.1cm}20$ GeV~\cite{5}.
As such, they
consistently fail to \mbox{reproduce} the experimental nucleus--nucleus
data.
Also,
attempts have been presented of \mbox{modifying} some of the existing models by implementing an IS-violating asymmetry between $u\bar{u}$ and $d\bar{d}$ production at the quark level. In order to describe the experimental data, an asymmetry of 3/1 in $u\bar{u}/d\bar{d}$ production is needed in the microscopic UrQMD model~\cite{b}. The microscopic SMASH requires the corresponding asymmetry of the order of 1.89~\cite{v}, while the simpler, semi-empirical QC model requires $u\bar{u}/d\bar{d}\hspace*{-0.1cm}=\hspace*{-0.1cm}1.2$~\cite{QC} which directly reflects the charged-over-neutral ratio $R_K$ reported by the experiment. While all of these works either provide or are expected to provide predictions for other particle production observables that can be verified experimentally, none of them gives an explicit reason for the size of the postulated asymmetry(ies) which, on the other hand, must remain very large at least for the cited microscopic models. Independently of the above, new ideas are proposed for IS violation by electromagnetic (EM) effects, phenomena related to finite density or the chiral anomaly, and others. A~compendium of these can be found in Ref.~\cite{5}. Within the limits of the knowledge of the author, these new ideas did not result as yet in quantitative theoretical predictions. Thus, the theoretical explanation of the reported unexpectedly large violation of isospin symmetry in multiparticle production is still to come.

A potentially interesting issue is addressed in the cited Ref.~\cite{b}. Assuming the very large, threefold asymmetry in $u\bar{u}$ over $d\bar{d}$ production mentioned above, the UrQMD calculation predicts asymmetries in charged-over-neutral kaon yields not only in nucleus--nucleus reactions as a function of energy, but also in $p$+$\hspace*{-0.02cm}p$ and $e$$^+$\hspace*{-0.06cm}+\hspace*{0.01cm}$e$$^-$ collisions at $\hspace*{-0.06cm}\sqrt{s}\hspace*{-0.06cm}=\hspace*{-0.06cm}12.3$~GeV and $\hspace*{-0.06cm}\sqrt{s}\hspace*{-0.06cm}=\hspace*{-0.06cm}3.050$~GeV, respectively. The calculation gives an overall fair description of the corresponding asymmetries seen in the experimental data from NA61/SHINE~\cite{k0spp} and BES~III~\cite{bes3}. On that basis, the authors claim that the reported effect of IS violation in the kaon sector is not specific to nuclear collisions, but remains a universal feature of particle production present already at the elementary ($e$$^+$\hspace*{-0.06cm}+\hspace*{0.01cm}$e$$^-$) level. Such claims need to be taken with caution
as both the $p$+$\hspace*{-0.02cm}p$ and the $e$$^+$\hspace*{-0.06cm}+\hspace*{0.01cm}$e$$^-$ reaction contain $known$ sources of IS violation (\mbox{initial} excess of $u$ quarks in the proton, QED component of the reaction).
%
%
The conclusions from Ref.~\cite{b}
can be confronted with these from the original BES~III paper, 
%
stating
that the experimental data at hadron energy $E_h\hspace*{-0.05cm}>\hspace*{-0.05cm}0.8$~GeV {support}
IS
in parton \mbox{fragmentation} processes~\cite{bes3}. Nevertheless,
%
%
the
issue
of the potential
universality of the observed effect
remains
%
%
of top importance, and may appear to be relatively easiest to
verify
experimentally.

\section{Plans of the NA61/SHINE Collaboration}
\label{plans}

While the potential impact of an unexpectedly large violation of IS on the understanding of particle production in strong interactions clearly calls for high precision measurements and a coordinated inter-experimental effort, NA61/SHINE can most efficiently contribute to the clarification of the system-size dependence of this effect as addressed in Sec.~\ref{theory} above. The plan proposed by the Collaboration includes studies of particle production in collisions of charge-symmetric ($N=Z$) nuclei:
 C+C, O+O, and Mg+Mg\hspace*{0.02cm}/\hspace*{0.02cm}B+B
 at energies $\sqrt{s_\mathrm{NN}}=5.1$, 7.6, and 16.8~GeV after the end of the CERN Long Shutdown 3~\cite{maja,bls3}. This is to be supplemented by measurements of kaon production in $\pi^+$+C and $\pi^-$+C collisions in a charge-symmetric ensemble~~\cite{a}. These are expected to clarify whether the unexpected effect seen in pion--carbon reactions (Sec.~\ref{small}) indeed corresponds to the violation of isospin symmetry, and what is its evolution from small ($\pi^\pm$+C)
systems up to intermediate nucleus--nucleus
 reactions. First, introductory data samples for both programs ($\pi^\pm$+C:~\hspace*{-0.05cm}$\sqrt{s_{\pi \mathrm{N}}}=17.3$~GeV, O+O:\hspace*{-0.05cm}~$\sqrt{s_\mathrm{NN}}=16.8$~GeV) have been recorded before the start of CERN LS3.

\section{Summary}
\label{sum}

The NA61/SHINE Collaboration reported the observation of an unexpectedly large violation of isospin symmetry in the kaon sector of multiparticle production in 0--10\% central Ar+Sc collisions at $\sqrt{s_\mathrm{NN}}=11.9$~GeV. This corresponds to an unexpected excess of charged-over-neutral kaon yields of $18\pm 6$\% at mid-rapidity and 4.2\,$\pm$\,1.6 additional charged kaons produced on the average per one Ar+Sc collision. Nucleus--nucleus results from other experiments in the energy regime $2.6<\sqrt{s_\mathrm{NN}}<200$~GeV and the preliminary NA61/SHINE Ar+Sc data at $\sqrt{s_\mathrm{NN}}=8.8$~GeV appear consistent with this observation. Known sources of IS violation contained in theoretical models do not explain this effect. At collision energies of $0.9<\sqrt{s_\mathrm{NN}}<13$~TeV, the charged-over-neutral kaon production ratio in $p$+$\hspace*{-0.02cm}p$, $p$+Pb and Pb+Pb reactions remains consistent with unity but within large experimental uncertainties, of the order of 10\%. 

NA61/SHINE data in $p$+$\hspace*{-0.02cm}p$ collisions at $7.7<\sqrt{s_\mathrm{NN}}<17.3$ GeV also exhibit a significant excess of charged-over-neutral kaon production, but the latter can be described in the framework of a semi-empirical model as resulting from the excess of $u$ quarks in the incoming protons. This is not the case for the corresponding charged kaon excess in $\pi^-$+C reactions at $\sqrt{s_\mathrm{NN}}=17.3$ and 25.7~GeV recorded by the same experiment, which goes against the predictions from several theoretical models. Further experimental studies of $\pi^+$+C and $\pi^-$+C collisions in a charge-symmetric ensemble are planned to elucidate whether the latter effect also corresponds to an unexpected violation of isospin symmetry. 

The experimental situation remains at the present moment (2026) with no clear explanation from theory. Some of the theoretical considerations suggest a close re-examination of kaon production in $e$$^+$\hspace*{-0.06cm}+\hspace*{0.01cm}$e$$^-$ collisions in view of the effects discussed above.

The potential high impact of an unexpected IS violation on the present understanding of multiparticle production calls for a coordinated theoretical and experimental effort, including precision measurements as a function of energy, theoretical predictions and experimental measurements for new observables, and studies of the system-size dependence of the reported effect. The latter issue can be addressed by NA61/SHINE in the regime of $\pi^\pm$+C reactions as well as collisions of light charge-symmetric nuclei up to $A=24$. 

A collection of experimental results and theoretical ideas on the above subject can be found in a White Paper~\cite{5} from a dedicated workshop organized in late 2025~\cite{ibr}. The next workshop in this series is planned around 19--24 April, 2027~\cite{ibr2}.

\vspace*{0.2cm}
{\bf Acknowledgments}

The author warmly thanks all the Organizers of a very interesting and versatile conference in the MESON series.
Also, many thanks to K.~Grebieszkow for her help in improving the manuscript.
This work was supported by the Polish Ministry of Science and Higher Education (contract No.~2025/WK/05).


\end{document}